TITLE:

**Seismic Wave Amplification in 3D Alluvial Basins:**

**3D/1D Amplification Ratios from Fast Multipole BEM Simulations**

AUTHORS' NAMES:

Kristel C. Meza-Fajardo[(1)(*)], Jean-François Semblat[(2)], Stéphanie Chaillat[(3)] and Luca Lenti[(4)]

CORRESPONDING AUTHOR:

Kristel C. Meza-Fajardo

Department of Civil Engineering,

Universidad Nacional Autónoma de Honduras

Tegucigalpa, Honduras

E-mail: kristelmeza@unah.edu.hn

(*) Formerly at IFSTTAR, Départ. GERS, 20 Boulevard Newton, Champs sur Marne, France.

**ABSTRACT**

In this work, we study seismic wave amplification in alluvial basins having 3D standard geometries through the Fast Multipole Boundary Element Method in the frequency domain. We investigate how much 3D amplification differs from the 1D (horizontal layering) case. Considering incident fields of plane harmonic waves, we examine the relationships between the amplification level and the most relevant physical parameters of the problem (impedance contrast, 3D aspect ratio, vertical and oblique incidence of plane waves). The FMBEM results show that the most important parameters for wave amplification are the impedance contrast and the so-called equivalent shape ratio. Using these two parameters, we derive simple rules to compute the fundamental frequency for various 3D basin shapes and the corresponding 3D/1D amplification factor for 5% damping. Effects on amplification due to 3D basin asymmetry are also studied and incorporated in the derived rules.

## INTRODUCTION

The amplification of seismic waves in alluvial deposits is strongly influenced by the geometry and mechanical properties of the surficial layers. The estimation of this amplification in codes is mainly performed through simplified 1D approaches. However the amplification process can significantly differ between the 1D (horizontal layering) and the 2D/3D cases because of focusing effects and waves generated at basin edges, (*e.g.*, Paolucci 1999). Some analytical and numerical results have been already derived by various authors for the response of basins with simple geometries to incident seismic waves. Bard and Bouchon (1985) studied rectangular and sine-shaped soft layers embedded in a rigid half space considering incident plane *SH*-waves. The propagation of plane vertical *SH*-waves in 2D cylindrical basins was analyzed by Semblat *et al.* (2010) and Bonnet (1999) using the Boundary Element Method in the frequency domain. Rodriguez-Zuñiga *et al.* (2005) studied the case of a 3D cylindrical basin having a rectangular vertical cross-section and found a large difference between the 2D and 3D response at the center of the basin. Papageorgiou and Pei (1998) considered incident body and Rayleigh waves in 3D cylindrical basins with semicircular cross-section. In the works of Bard and Bouchon (1985) and Jiang and Kuribayashi (1988), it was reported that the fundamental frequencies of the basins only depend on the aspect ratio and the 1D fundamental frequency at the center of the valley. The 3D wave diffraction by a semi-spherical canyon has been also studied (Lee 1978; Kim and Papageorgiou 1993; Yokoi 2003; Liao *et al.* 2004; Chaillat *et al.* 2008) and 3D wave amplification due to surface heterogeneities has also been quantified (Sánchez-Sesma and Luzón, 1995; Komatitsch and Vilottte 1998; Drawinski 2003; Moczo *et al.* 2002; Chaillat *et al.* 2009). Smerzini *et al.* (2011) made comparisons of 3D, 2D and 1D amplification using the Spectral Element Method with a 3D model of the Gubbio plain in Italy. Olsen *et al.* (2000) found differences among 3D/2.5D/1D amplification and duration with a 3D finite difference model of the Upper Borrego Valley, California. The 2D amplification features may be interpreted through 2D/1D amplification factors with respect to the case of a horizontal layer (Chavez-Garcia and Faccioli 2000, Gélis *et al.* 2008; Makra *et al.* 2005; Semblat *et al.* 2010). The estimation of 3D/1D amplification factors in different 3D configurations is the main goal of this paper.

The 3D amplification of seismic waves is modeled through the Fast Multipole Method (FMM). This formulation of the Boundary Element Method (BEM) allows the acceleration of iterative solvers for the global linear system of equations. The application of the Fast Multipole Boundary Element Method (FMBEM) is beneficial in problems of elastic wave propagation involving strong velocity gradients or 3D unbounded domains since large BEM meshes are required (Makra *et al.* 2005; Chaillat *et al.* 2008; Chaillat *et al.* 2009; Delépine & Semblat 2012). Extension of the FMBEM to propagation in weakly dissipative viscoelastic media was carried out by Grasso *et al.* (2012). In this work, we apply the formulation of the FMBEM for viscoelastic media to study wave amplification phenomena in 3D basins with canonical geometries subjected to incident plane waves. Following the work done by Makra *et al.* (2005) and Semblat *et al.* (2010) for canonical or realistic 2D configurations, we identify the fundamental frequencies of the basin, and study the relationships between the amplification level and relevant mechanical parameters such as impedance contrast, aspect ratio and damping. Finally, we synthesize the results to propose simple rules for 3D/1D amplification factors including 3D basin asymmetry effects.

## THE FAST MULTIPOLE BOUNDARY ELEMENT METHOD

Following the discussion in Grasso *et al.* (2012), the classical boundary integral representation formula gives the displacement $u$ in the direction $k$ at point $\mathbf{x}$ in the interior $\Omega$ of an isotropic homogeneous viscoelastic solid in the absence of body forces as:

$$u_k(\mathbf{x}) = \int_{\partial\Omega} \left[ t_i(\mathbf{y}) U_i^k(\mathbf{y}-\mathbf{x}; \boldsymbol{\omega}) - u_i^k(\mathbf{y}) T_i^k(\mathbf{x}, \mathbf{y}; \boldsymbol{\omega}) \right] dS_y \tag{1}$$

where $\mathbf{t}$ is the traction vector, $U_i^k(\mathbf{y}-\mathbf{x}; \omega)$ and $T_i^k(\mathbf{x}, \mathbf{y}; \omega)$ denote respectively the *i*-th component of the displacement and traction vectors associated to the viscoelastic fundamental solution generated at a point $\mathbf{y}$ by a unit point force applied at $\mathbf{x}$ along the direction $k$, given by:

$$U_i^k(y-x; \omega) = \frac{1}{\mu \hat{k}_s^2} \left[ \left( \delta_{qs}\delta_{ik} - \delta_{qk}\delta_{is} \right) \frac{\partial}{\partial x_q} \frac{\partial}{\partial y_s} G\left( \|\mathbf{y}-\mathbf{x}\|; \hat{k}_s^2 \right) + \frac{\partial}{\partial x_i} \frac{\partial}{\partial y_k} G\left( \|\mathbf{y}-\mathbf{x}\|; \hat{k}_p^2 \right) \right]$$

$$T_i^k(\mathbf{x},\mathbf{y};\omega) = \hat{C}_{ijhl}(\omega) \frac{\partial}{\partial y_l} U_h^k(\mathbf{y}-\mathbf{x}; \omega) n_j(\mathbf{y}) \tag{2}$$

where $\hat{k}_{p,s}(\omega) = k_{p,s}(\omega)[1 + \beta(\omega)]$, $\mathbf{n}(\mathbf{y})$ is the outward unit normal, $G(r; \hat{k})$ is the free-space Green's function for the Helmholtz equation, given by:

$$G(r; \hat{k}) = \frac{\exp(i\,\hat{k}r)}{4\pi r} = \exp(-\beta k r) \frac{\exp(ikr)}{4\pi r} \quad \beta > 0 \tag{3}$$

$\beta(\omega)$ is the material damping ratio, and $\hat{C}_{ijhl}$ is given by the constitutive relation for a viscoelastic medium in terms of the complex valued Lamé parameters $\hat{\lambda}(\omega)$ and $\hat{\mu}(\omega)$:

$$\hat{C}_{ijhl}(\omega) = \hat{\lambda}(\omega)\delta_{ij}\delta_{kl} + \hat{\mu}(\omega)\delta_{ij}\delta_{kl}\left( \delta_{ik}\delta_{jl} + \delta_{il}\delta_{jk} \right) \tag{4}$$

When the boundary conditions $t^L$ are imposed, the integral representation of eq. (1) leads to:

$$(Ku)(\mathbf{x}) = f(\mathbf{x}) \quad (\mathbf{x} \in \partial\Omega) \tag{5}$$

where:

$$(Ku)(\mathbf{x}) = c_{ik}(\mathbf{x})u_{ik}(\mathbf{x}) + (P.V.) \int_{\partial\Omega} u_i(\mathbf{y}) \, T_i^k(\mathbf{x}, \mathbf{y}; \omega) dS_y$$

$$f(\mathbf{x}) = \int_{\partial\Omega} t_i^L(\mathbf{y}) U_i^k(\mathbf{x}, \mathbf{y}; \omega) \, dS_y \tag{6}$$

where (P.V.) indicates the Cauchy principal value singular integral. The free term $c_{ik}$ depends on the local geometry of the boundary and is equal to $\delta_{ik}/2$ for any point on a smooth boundary.

Once Eq. (5) has been discretized using $N_I$ isoparametric boundary elements, a square complex-valued matrix equation of size $N = 3N_I$ is obtained. The coefficient matrix is fully populated, making its storage and resolution impractical for models with sizes exceeding $N = O(10^4)$. The FMBEM tackles this problem by reformulating and expanding the fundamental solutions in terms of products of functions of **x** and **y**, lowering the overall complexity to $O(NlogN)$ (for the multilevel formulation see Chaillat *et al*., 2008). The expansion of the fundamental solution needs to be truncated by computational necessity and the truncation parameter $m$ needs to be carefully chosen as to guarantee sufficient accuracy. To properly choose the truncation parameter $m$, empirical criteria has been developed for elastodynamics by Chaillat *et al*. (2008). For the viscoelastic case, Grasso *et al*. (2012) established an optimum rule to select $m$ which is linearly related to the damping ratio $\beta$.

## NUMERICAL MODELS FOR VARIOUS 3D BASIN GEOMETRIES

The physical model we use for this study corresponds to a basin embedded in a halfspace, which is subjected to an incident field of plane *P* or *S* waves, as shown in Figure 1. Since the FMBEM used here is based on the full-space fundamental solution, the mesh must include the free surface which, for implementation purposes, must be truncated as shown in Figure 1. According to Jiang and Kuribayashi (1988) and Grasso *et al*. (2012), a truncation radius *L*=5*R* where *R* is the radius of the basin at the free surface, gives satisfactory results.

The interface between the basin and the underlying bedrock follows the equation:

$$\left(\frac{x}{R}\right)^2 + \left(\frac{y}{R}\right)^2 + \left(\frac{z}{h}\right)^2 = 1, \qquad z \leq 0 \tag{7}$$

where $h$ is the maximum depth of the valley. The geometrical and mechanical properties of the problem are described in Table 1, where $v_{s1}$ and $v_{s2}$ are the shear wave velocities of the half space and the basin, respectively.

Several values of the horizontal aspect ratio are considered $\kappa_h = R/h = 0.5, 1, 2, 3$, ranging from deep to shallow basins, as shown in Fig. 2. Different velocity ratios are also considered, $\chi = 2, 3, 4, 6$, the higher the value of $\chi = v_{s1}/v_{s2}$, the softer the basin. We also consider incident *P* and *S* waves, at incidence angles $\theta = 45°, 30°$ and $0°$ (vertical incidence). The size of the resulting numerical models ranged from 16461 to 92565 degrees of freedom.

## BASIN AMPLIFICATION WITH ELASTIC MODEL

### *Previous results on 2D amplification*

Previous researches on 2D canonical basins give the variations of the amplification level and the related frequencies as parametric functions of the horizontal aspect ratio $\kappa_h = R/h$ and the velocity ratio $\chi = v_{s1}/v_{s2}$, (Semblat *et al*. 2010). These numerical results for elliptical basins (Fig. 2) have been computed with the classical BEM and are in agreement with Bard and Bouchon's results (1985) for sinusoidal basins in terms of fundamental frequencies, that is:

$$f_{2D} = \frac{v_{s2}}{4H}\sqrt{1 + \kappa_v^2} \tag{8}$$

where $\kappa_v = 1/\kappa_h = h/R$ is the vertical shape ratio, $R$ is the basin half width, $h$ is the basin depth, and $v_{s2}$ is the velocity in the basin.

As shown in Fig. 3, the 2D basin amplification in the elastic case may be very large (up to 50!) when compared to the 1D case (dotted line). It shows the influence of 2D basin effects in the undamped case. For realistic basin geometries and damped media, Makra *et al.* (2005) and Gélis *et al.* (2008) proposed 2D/1D amplification factors around 3 in a given frequency range. In the following, the 3D/1D amplification factor will be assessed in both the undamped and damped cases.

*P-wave amplification in 3D elastic basins*

The amplification due to wave diffraction caused by the presence of an elastic semi-spherical basin in the propagation media has been already studied by Sánchez-Sesma (1983) for the case of vertically incident *P*-waves. In his study Sánchez-Sesma reports an amplification level of 170 per cent (relative to the amplitude of the free field) for the particular case of a basin with aspect ratio $R/h = 1$ and velocity ratio $\chi = 1.414$ at a frequency of about 0.23 Hz. The need to study the effects of interface depth and curvature on wave amplification was already emphasized by Sánchez-Sesma (1983). In Chaillat *et al*. (2009), the accuracy of the FM-BEM results have been verified by comparing them with solutions provided by Sanchez-Sesma, (1983).

In this work, we define the amplification factor as the maximum of the displacement at the free surface of the basin $U^B$ over the maximum of the displacement at the free surface of the half space. Since incident waves are reflected at the free surface of the half space, the total displacement field at the free surface of the half space $U^F$ is composed of the incident and reflected waves. This total displacement field $U^F$ is the input of the numerical model, which is imposed at the boundary limiting the half space. The amplification can then be expressed as:

$$A = \frac{\max(U^B)}{\max(U^F)} \qquad (9)$$

It is important to note that we take the value $\max(U^B)$ regardless of its place of occurrence, and therefore, its location might or might not be the center of the basin. However, we know that for the fundamental mode of basin vibration, the maximum displacement will occur at the center of symmetrical basins both for *P* and *S* waves. It is in the case of the higher modes of vibration that the location of maximum displacement may not be at the center of the basin and such location will be different for various frequencies (see for instance, plots of vibrations modes in Jiang and Kuribayashi, 1988). This fact should be taken into account when interpreting amplification for higher modes with Eq. (9). In Fig. 4 we can observe the 3D amplification factor for the vertical component due to a vertically incident *P*-wave when we fix the aspect ratios $R/h$ and vary the ratio of velocities $\chi$. The results are plotted as a function of the normalized frequency $\tilde{f} = f/f_{rp}$, where $f_{rp} = v_{p2}/4h$ is the 1D resonance frequency at the center of the basin, and $v_{p2}$ is the *P* wave velocity of the basin. Let us note that the velocity $v_{p2}$, needed to compute $f_{rp}$ can be computed as $v_{p2} = v_{s2} \cdot \gamma$, where $v_{s2} = v_{s1}/\chi$ and $\gamma = \sqrt{(1 - 2v_2)/[2(1 - v_2)]}$. We take $v_{s1} = 1$ km/s, and the Poisson ratio of the basin $v_2$ and its height $h$, are given in Table 1. Then for a basin with $\chi = 2$ the 1D resonance frequency is $f_{rp} = 0.234$ Hz. In figure 4 the reported maximum normalized frequency is 4, therefore the maximum frequency considered for this basin is $(0.234 \times 4) = 0.94$ Hz.

We can observe in Figure 4 that the aspect ratio $R/h$ mainly affects the fundamental frequency. The higher the aspect ratio (that is, the shallower the basin), the lower the fundamental frequency. For the case $\chi = 6$ (Fig. 4b) the amplification can reach unrealistic values of up to 90. We can confirm that in most cases, the ratio of velocities does not affect the fundamental frequency of the 3D basin. We can also observe in Figure 4 that for some basins with aspect ratio $R/h = 3$, the 3D peak frequency appear lower than the 1D frequency. Because these are very shallow basins, the peak is dominated by the effects of the "wave trapping" and not by the fundamental modes of vibration as discussed by Bard and Bouchon (1985) in the 2D case.

In Table 2 and Fig. 5 we can observe the amplification spectra for the fundamental and second modes of a semi-spherical basin subjected to incident plane $P$-waves. By amplification spectrum we mean the plot showing the peak values of the amplification function (9) versus the frequencies at which those peaks occur. The plots in Figure 5 basically summarize the two first peaks of the amplification functions shown in Figure 4. For each value of the velocity ratio $\chi$ the plots move from lower to higher frequencies as the shape ratio $R/h$ decreases. The variation of peak amplification with $R/h$ however, is not significant. The effects of the aspect ratio and the ratio of velocities are similar to that derived by Semblat *et al.* (2010) in the 2D case, but the level of amplification considering the 3D effects is much higher. However, in realistic cases, there will be attenuation due to material damping and therefore the amplification levels will not be as high as those found for this undamped model.

## BASIN AMPLIFICATION WITH 3D DAMPED MODEL

*Incident P-wave*

Considering a 3D basin model with 5% damping and a halfspace with 0.5% damping, the amplification levels for the vertical component for a vertically incident $P$-wave are shown in Fig. 6 for several aspect $R/h$ and velocity ratios $\chi$. The results show there is no significant change in the resonance frequency with damping, but the amplification factor is significantly reduced, with a stronger reduction for the second and higher modes. Amplification factors for the 3D damped semi-spherical basin subjected to vertically incident $P$-waves are shown in Table 3 and Fig. 7. The higher amplification level observed in the figure is around 14 and corresponds to the deepest and the softest basin. For the basin with $\chi = 2$ on the other hand, the amplification factor gets down to 3.

Now, we consider different directions of incidence for the incoming field. In Fig 8, we compare the results for $\theta = 0°$ (vertical incidence) and $\theta = 30°, 45°$, for the case of the deepest 3D basin. We notice that the 3D amplification of the vertical component at the fundamental and predominant frequencies decreases with non-vertical incidence.

*Incident S-wave*

In this part we study the 3D amplification due to an incident $S$-wave. The $S$-wave induces important amplification in two components of displacement, the vertical component, and the horizontal component oriented in the direction of the $S$-wave polarization (in this example the direction of the $x$-axis). Thus, we compute the amplification factors $A_H$ and $A_V$, for the horizontal and vertical displacement components respectively, using the total magnitude of the free field displacement $U^F$ as follows:

$$A_V = \frac{\max(U_z^B)}{\max(U^F)}, \qquad A_H = \frac{\max(U_H^B)}{\max(U^F)} \tag{10}$$

where $U_z^B$ and $U_H^B$ are the magnitudes of the horizontal and vertical components of displacement at the top of the basin, respectively. Figure 9 illustrates amplification factors for *S*-waves with incidence angles $\theta = 0°, 30°, 45°$ for the deeper 3D basins ($R/h = 0.5$, $R/h = 1$). In this case the frequency is normalized as $\tilde{f} = f/f_{rs}$, where $f_{rs} = v_{s2}/4h$. In order to compare how the amplification factors $A_V$ and $A_H$ vary with $\tilde{f}$, we have plotted $-A_V$ in the same axis as $A_H$, with the understanding that they both are always positive, as indicated by definitions (10). As expected, as the incidence angle increases, the amplification of the horizontal component is reduced, whereas the amplification of the vertical component is increased. It is important to note that surface waves contribute to the higher amplification in the vertical component when $\theta = 45°$.

**EFFECTS OF 3D BASIN SHAPE**

In this section we study the influence of different simple geometries on wave amplification in 3D alluvial basins. Results for 2D effects have been derived by Sanchez-Sesma and Velazquez (1987) for an infinite dipping layer, and by Paolucci and Morstabilini (2006) for a wall-layer system. A soil layer of thickness $h$, delimited laterally by a rigid bedrock dipping vertically is referred as a wall-layer system. Paolucci and Mostabilini (2006) obtained 2D/1D amplification between 1.4 and 1.7 considering 2.5% and 1% damping factors for the soil layer. The 2D amplification factors estimated in these studies had a maximum of 1.7 for the most severe case. Also for the 2D case, amplification in rectangular and sine-shaped basins was investigated by Bard and Bouchon (1985), concluding that 2D resonance dominates in basins with large aspect ratio. An amplification factor of about 1.75 was reported by Bonnet (1999) for 2D cylindrical basins with circular cross-section, at the center of the basin.

We compare here the amplification factor for three different 3-D shapes. The elliptical shape has already been described in the previous sections and in equation (7). The second shape we consider corresponds to a super-ellipsoid of fifth degree, given by the following equation:

$$\left|\frac{x}{R}\right|^5 + \left|\frac{y}{R}\right|^5 + \left|\frac{z}{h}\right|^5 = 1, \qquad z \leq 0 \tag{11}$$

When the exponent of the super-ellipsoid is 5 the basin geometry is closer to a "box" shape as shown in Figure 10. The third shape we consider in this study is a 3D cosine shape (see Fig. 11), which can be expressed as:

$$4p^2 \left(\frac{z}{h}\right) = \left[\cos\left(\pi s \frac{x}{R}\right) + 2p - 1\right]\left[\cos\left(\pi s \frac{y}{R}\right) + 2p - 1\right] \tag{12}$$

where $p$ is the percentage of the height of a full cosine cycle (in this study $p = 0.9$) and the parameter $s$ is given by:

$$\cos(\pi s) = 1 - 2p \tag{13}$$

Note that the total depth of the basin is 1, but the shape is scaled such that its height is only 90% of the total height of a full cosine cycle. We selected this percentage instead of the complete height of the cosine (for which $p$ would be 1) to avoid numerical artifacts at the intersection of the basin boundary and the free surface, which occur when two surfaces intersect with the same tangent.

Considering the "equivalent shape ratio" $\tilde{\kappa}_h = l_0/h$ used by Jiang and Kirubashi (1988), where $l_0$ is the half width over which the depth of the basin is half its maximum value, we can take into account the difference in thickness of the three basin shapes, as shown in Figure 12. We can see in Figure 12(b) that even though the three shapes have the same aspect ratio ($\kappa_h = R/h$) the cosine shape has the lowest equivalent shape ratio.

In Figure 13 we can see the effect of different equivalent shape ratios on the amplification factors for the three different basin shapes, when they are subjected to vertically propagating $S$-waves, for the case of an aspect ratio $\kappa_h = 0.5$ (Figures 13a-13c) and $\kappa_h = 2$ (Figures 13e-13f). The amplification level for the ellipsoidal shapes remains almost the same, both for the horizontal and vertical components; however the difference in the exponent $n$ leads to different fundamental and dominant frequencies. On the other hand, we can see slightly higher amplification factors for the fundamental frequency in the case of the basin with cosine shape, a result that was expected, since there are strong basin edge effects and there is more trapping of waves due to the lower thickness of the basin. The same trends are observed in Figure 14, which shows the amplification on top of the three different basins for S-waves with 30° angle of incidence. There is higher amplification for the vertical component for the cosine and ellipsoidal basins when $R/h = 2$ and $\theta = 30°$, however, it corresponds to higher modes. In Figure 15, we compare the fundamental frequency obtained for the three different basin shapes for the aspect ratios $R/h = 0.5$ (Fig. 15a) and $R/h = 2$ (Fig. 15b). We can observe that in both cases the normalized fundamental frequency $f/f_{rs}$ has little variation with the velocity ratio (below 25% for $R/h = 0.5$ and about 7% for $R/h = 2$) and is in fact mainly determined by the equivalent shape ratio. For the basin with aspect ratio $R/h = 0.5$ the maximum variation of the normalized fundamental frequency with the equivalent aspect ratio is of 44%. In the case of the normalized predominant frequency $f/f_{ps}$ (second mode), shown in Figure 16, we have similar conclusions. According to Figure 16, the maximum variation of the normalized predominant frequency is of 7% with the velocity ratio, and of 45% with the equivalent shape ratio.

## EFFECTS OF ASYMMETRY

In this section we assess the effects of basin asymmetry, changing the radius of the basin in the $x$- or $y$-direction. In Figure (17a) the radius in the $y$-direction is twice the radius in the $x$-direction, and in Figure (17b) the radius in the $x$-direction is twice the radius in the $y$-direction. Thus, the parameter $\kappa_{xy} = R_x/R_y$ will be different for these two basins. The effect of asymmetry will then be different for these two cases since the dimension of lower basin thickness would be either parallel or perpendicular to the direction of polarization of the incident $S$-wave. We also consider the case when the direction of polarization of the incident plane waves does not coincide with the direction of any basin radii but it is 45° from the $x$-axis [Figure (17c)].

In Figure 18 and in Figure 19 we can see the amplification factor for the symmetric, and the two asymmetric basins, for the aspect ratios $R/h = 0.5$ and $R/h = 3$, respectively. It is observed in these two cases that the basin response is similar for the symmetric and the asymmetric basin with $\kappa_{xy} = 2$. This is related to the fact that these basins have the same radius in the direction of polarization of the incident wave ($x$- direction). Whereas the asymmetric basin with $\kappa_{xy} = 0.5$ the radius in the direction of polarization is half of the other two, and thus, its response differs even though the amplification level is only slightly different. Figures 18(d) and 19(d) show that when the direction of polarization is 45° from the $x$- axis the amplification is reduced. Another way to characterize the effects of asymmetry and the direction of polarization of the incident wave is with the product $\kappa_R\left(l_p/R\right)$ where $\kappa_R$ is the ratio between the larger radius $R_L$ and shorter radius $R$ of the basin, and $l_p$ is half the dimension of the basin parallel to the direction of polarization of the incident wave. Note that for a symmetric (semi-spherical) basin $\kappa_R = 1$ and $l_p/R = 1$. In Figure 20 we present the results for the asymmetric basins in the form of isovalue curves of amplification as function of the parameter $\kappa_R\left(l_p/R\right)$ and normalized frequency. From the results we conclude that the main effect of basin asymmetry is an increase in the fundamental and predominant frequencies. The lowest fundamental frequency in Fig. (20) corresponds to the symmetric basin, that is, to the case $\kappa_R\left(l_p/R\right) = 1$.

Figure 21 shows the normalized fundamental frequency for the asymmetric basins when subjected to vertically incident *S*-waves. In this figure the basin fundamental frequency is normalized with respect to the frequency of the symmetric basin. As expected, we can observe in Figure (21) that as the aspect ratio increases (shallower basins), the difference in fundamental frequency due to asymmetry is reduced. Furthermore, as it was observed for the symmetric case, the fundamental frequencies are practically independent of the velocity ratio $\chi$. On the other hand, Figure (22) shows amplification in asymmetric basins with respect to amplification on symmetric ones. The results show that basin asymmetry leads to higher amplification levels for the case $R_x/R_y = 2$ (an increase of about 10% with respect to symmetric basins), the difference being more pronounced (with a reduction of up to 50%) for the narrower basins with lower velocity contrast $\chi$.

## SIMPLE RULES TO ASSESS 3D BASIN EFFECTS

Based on the results presented in the previous sections we derive simple rules to compute the fundamental frequencies and the associated amplification factors, in terms of two parameters, the equivalent shape ratio $\tilde{\kappa}_h = l_0/h$ and the factor $\tilde{\chi} = (\xi_2/2\xi_1)(\rho_1/\rho_2)\chi$, where $\xi_2 = 5\%$ and $\xi_1 = 0.5\%$ are the damping ratios of the basin and the half space, respectively . We are particularly interested in the case of incident *S*-waves. Since in this section we propose rules for amplifications corresponding to the fundamental mode of vibration, it should be understood that its location is the center of the basin. Now, for the fundamental frequency of the basin we propose the following relation:

$$\frac{f_0}{f_{rs}} = 1 + (\tilde{\kappa}_h)^{-1.24} \tag{14}$$

The plot of the proposed equation along with the data used for its derivation is presented in Figure 23. We also plot in Figure 23 the rule to compute the normalized fundamental frequency for shear response proposed by Jiang and Kuribayashi (1988). The two rules predict practically the same results. Note that when the equivalent shape ratio $\tilde{\kappa}_h$ becomes large (shallower basins), the fundamental frequency tends to the 1D resonance frequency. On the other hand, Figure 24 shows the amplification on top of the basin corresponding to the fundamental frequency. We obtain the 3D/1D amplification factor when we normalize this 3D amplification with the factor $\tilde{\chi} = (\xi_2/4\xi_1)(\rho_1/\rho_2)\chi$, consisting of the elastic 1D resonance amplification $(\rho_1/\rho_2)\chi$, and a damping relation $(\xi_2/4\xi_1)$. The data obtained from our simulations (normalized $\tilde{\chi}$) is very close to the 3D/1D amplification factors computed by Olsen (2000) for long period (0-0.5 Hz) basin response using 3D finite difference simulations for several southern California basins. He reported 3D/1D velocity amplification factors less than about 4 at the deepest parts of the basins. Now, using the results from our simulations (shown in Figure 24), we propose to define the 3D/1D amplification factor in three regions as follows:

$$\frac{A}{\tilde{\chi}} = \begin{cases} 1.7 + 1.42\tilde{\kappa}_h & \tilde{\kappa}_h \leq 1.2 \\ 3.40 & 1.2 < \tilde{\kappa}_h \leq 2 \\ 1 + 3.64(\tilde{\kappa}_h)^{-0.60} & 2 < \tilde{\kappa}_h \end{cases} \tag{15}$$

The predicted 3D/1D amplification factor obtained with this equation is also shown in Figure 24. Even though Jiang and Kuribayashi (1988) also proposed a rule to compute basin amplification, we do not compare it here with ours since it is not formulated in terms of 3D/1D response. Besides, the rule by Jiang and Kuribayashi (1988) is based on computations considering damping only for the basin (elastic halfspace), whereas Eq. (15) is based on computations we performed with 0.5% damping for the halfspace. Thus, because of the difference made by the radiation of energy at the halfspace, the predictions made with the rule of Jiang and Kuribayashi (1988) will be higher than those obtained with Eq. (15). For values of the equivalent aspect ratio greater than two, the proposed function is decreasing, and when $l_0/h$ approaches infinity, the 3D/1D amplification factor tends to that of the 1D elastic response times the damping relation. Finally, these rules to estimate the 3D fundamental frequency and 3D/1D amplification factor should also be corrected for basin asymmetry effects. We propose the following linear rule to correct the fundamental frequency for such effects:

$$\frac{f_{0(asym)}}{f_{0(sym)}} = 1.36 - 0.1\kappa_h \tag{16}$$

where $f_{0(asym)}$ and $f_{0(sym)}$ are the fundamental frequencies of the asymmetric and symmetric basin, respectively. The parameter $\kappa_h$ is the aspect ratio obtained using the lower radius, and the parameter $l_0$ in Equation (14) corresponds to the lower half-width of the asymmetric basin. Now, since we propose Eq. (16) after observing from our computations an increase in the fundamental frequency due to asymmetry, then the relation $f_{0(asym)}/f_{0(sym)}$ given by (16) should be greater than 1. Thus, Eq. (16) is valid only for shape ratios $\kappa_h \leq 3.6$. Finally, for the 3D/1D amplification factor $A_{(asym)}$ of the asymmetric basin we suggest a simple increase of 10% in the amplification $A_{(sym)}$ of the symmetric basin:

$$A_{(asym)} = 1.1A_{(sym)} \tag{17}$$

With the application of formula (17), the maximum amplification factor predicted with our rules is of 3.74.

**CONCLUSIONS**

We have computed the amplification of seismic waves in 3D basins when subjected to incident fields of plane $P$ and $S$ waves through the Fast Multipole Boundary Element Method in the frequency domain. We considered basin models with and without attenuation due to material damping. In accordance with previous studies, the FMBEM results show that the largest amplification levels correspond to the deepest basin with the strongest impedance ratio. However, we have found the ratio of velocities and the asymmetry to be the most important parameters for 3D wave amplification. The effects of asymmetry appear to be more significant for the fundamental frequency as the level of amplification was found to be increased by only 10%. In this study we considered 5% basin damping, and we found amplification factors at the top of the basin due to incident body waves to be close to four times higher than the 1D amplification level. Simple rules were derived to compute the fundamental frequency and its corresponding amplification factor at the center of 3D alluvial basins, with respect to the 1D elastic response. The proposed equations are expressed in terms of only the equivalent shape ratio $\tilde{\kappa}_h$, the basin/bedrock impedance contrast and the damping ratios.

As discussed in Semblat *et al.* (2005) the combined effect of basin geometry and soil layering may be important for amplification, and it should be considered for future, more detailed studies. Besides, the results obtained in this investigation are limited to low frequencies and small deformations that fall within the linear approximation of the basin response. However, simple criteria and practical rules should be targeted in order to make the results useful for practitioners.

**DATA AND RESOURCES**

All computations made in this work were performed using COFFEE, a FORTRAN code based on the FMBEM, developed by Stéphanie Chaillat for frequency-domain elastodynamics, and extended to viscoelasticity by Eva Grasso. The 3D meshes were generated with the GMSH code developed by C. Geuzaine and J.F. Remacle.

**ACKNOWLEDGEMENTS**

This research was financed by Electricité de France (EDF) through the MARS project (Méthodes Avancées pour le Risque Sismique).

**AUTHORS' AFFILIATIONS**:

(1)   Universidad Nacional Autónoma de Honduras, Tegucigalpa, Honduras. E-mail: kristelmeza@unah.edu.hn

(2)   Université Paris-Est, IFSTTAR, Départ. GERS, 20 Boulevard Newton, Champs sur Marne, France. E-mail: jean-francois.semblat@ifsttar.fr

(3)   POems (UMR 7231 CNRS), Applied Mathematics Department, ENSTA, Paris, France. E-mail: stephanie.chaillat@ensta.fr

(4)   Université Paris-Est, IFSTTAR, Départ. GERS, 20 Boulevard Newton, Champs sur Marne, France. E-mail: luca.lenti@ifsttar.fr

<u>FIGURE CAPTIONS</u>

Figure 1. A 3D semi-cylindrical basin embedded in a half space and subjected to a plane wavefield.

Figure 2. Basins with different aspect ratios. (a) $R/h = 0.5$ (b) $R/h = 1$ (c) $R/h = 2$ (d) $R/h = 3$.

Figure 3. Maximum amplification and related frequencies (in Hz) for variable shape ratios $kh$ and velocity ratios $\chi$. Here the symbol L has been used instead of R for the radius of the basin. Taken from Semblat *et al*. (2010).

Figure 4. Amplification at the top of the undamped 3D basin due to vertically incident *P*-waves for different aspect ratios. (a) $R/h = 0.5$ (b) $R/h = 1$ (c) $R/h = 2$ (d) $R/h = 3$.

Figure 5. Amplification spectra for the first and second elastic modes of a semi-spherical basin for a vertically incident *P*-wave. (a) First mode (b) Second mode.

Figure 6. Amplification at the top of the 3D basin with 5% damping due to vertically incident *P*-waves. (a) $R/h = 0.5$ (b) $R/h = 1$ (c) $R/h = 2$ (d) $R/h = 3$.

Figure 7. Amplification spectra for the first and second modes of 3D damped semi-spherical basin due to a vertically incident *P*-wave. (a) First mode (b) Second mode.

Figure 8. Amplification at the top of the 3D deepest basin with 5% damping and $R/h = 0.5$ due to incident *P*-waves. (a) Vertical incident (b) $\theta = 30$ degrees (c) $\theta = 45$ degrees

Figure 9. Comparison of amplification at the top of the 3D basin with 5% damping due to *S*-waves with different angles of incidence. (a) $R/h = 0.5$, $\theta = 0$ degrees (b) $R/h = 0.5$, $\theta = 30$ degrees (c) $R/h = 0.5$, $\theta = 45$ degrees (d) $R/h = 1$, $\theta = 0$ degrees (e) $R/h = 1$, $\theta = 30$ degrees (f) $R/h = 1$, $\theta = 45$ degrees.

Figure 10. 3D basin model with super-ellipsoid shape. (a) View of the mesh for the super-ellipsoid and the truncated free surface. (b) A cross section corresponding to the $x$-$z$ plane.

Figure 11. 3D basin model with shape of a 3D half sine. (a) View of the mesh for the 3D cosine and the truncated free surface. (b) A cross section corresponding to the $x$-$z$ plane (the period is adjusted so that the height of the bell is only 90% of that of a full cosine cycle).

Figure 12.Basin shapes considered in the study. (a) Cross section corresponding to the $x$-$z$ plane. (b) Equivalent shape ratio $l0/h$ for the three basin shapes for a unitary radius.

Figure 13. Amplification at the top of the 3D basin with 5% damping due to vertically incident *S*-waves with different equivalent shape ratios. (a) $R/h = 0.5$, basin with cosine shape (b) $R/h = 0.5$, basin with ellipsoidal shape, $n = 2$ (c) $R/h = 0.5$, basin with super-ellipsoidal shape, $n = 5$ (d) $R/h = 2$, basin with cosine shape (e) $R/h = 2$, basin with ellipsoidal shape, $n = 2$ (f) $R/h = 2$, basin with super-ellipsoidal shape, $n = 5$.

Figure 14. Amplification at the top of the 3D basin with 5% damping due to $S$-waves with incidence angle $\theta = 30°$. (a) $R/h = 0.5$, basin with cosine shape (b) $R/h = 0.5$, basin with ellipsoidal shape, $n = 2$ (c) $R/h = 0.5$, basin with super-ellipsoidal shape, $n = 5$ (d) $R/h = 2$, basin with cosine shape (e) $R/h = 2$, basin with ellipsoidal shape, $n = 2$ (f) $R/h = 2$, basin with super-ellipsoidal shape, $n = 5$.

Figure 15. Normalized fundamental frequency $f0/frs$ as a function of equivalent shape ratio $kh$. (a) Basin with aspect ratio $\kappa h = 0.5$ (b) Basin with aspect ratio $\kappa h = 2$.

Figure 16. Normalized predominant frequency $fp/frs$ as a function of equivalent shape ratio $\kappa h = 0.5$ (a) Basin with aspect ratio $\kappa h = 0.5$ (b) Basin with aspect ratio $\kappa h = 2$.

Figure 17. 3D basin models with asymmetry. (a) Basin with $Rx/Ry = 0.5$ (b) Basin with $Rx/Ry = 2$. (c) Basin with $Rx/Ry = 2$, direction of polarization of incident plane wave 45°.

Figure 18. Amplification at the top of the 3D basin with 5% damping due to vertically incident $S$-waves for $R/h = 0.5$. (a) Symmetric basin (b) Basin with $\kappa xy = 0.5$ (c) Basin with $\kappa xy = 2$ (d) Basin with $\kappa xy = 2$ and polarization direction at 45°.

Figure 19. Amplification at the top of the 3D basin with 5% damping due to vertically incident $S$-waves for $R/h = 3$. (a) Symmetric basin (b) Basin with $\kappa xy = 0.5$ (c) Basin with $\kappa xy = 2$ (d) Basin with $\kappa xy = 2$ and polarization direction at 45°.

Figure 20. Amplification factor at the top of the basin with 5% damping due to vertically incident $S$-waves for 3D symmetric and asymmetric basins.

Figure 21. Fundamental frequency in a 3D asymmetric basin normalized with respect to the fundamental frequency of the symmetric basin. (a) Basin with $Rx/Ry = 0.5$ (b) Basin with $Rx/Ry = 2$.

Figure 22. Amplification in a 3D asymmetric basin normalized with respect to the amplification of a symmetric basin. (a) Basin with $Rx/Ry = 0.5$ (b) Basin with $Rx/Ry = 2$.

Figure 23. Normalized fundamental frequency $fofrs$ as a function of equivalent shape ratio.

Figure 24. 3D/1D amplification factor $A/[(\xi2/4\xi1)(\rho1/\rho2\chi)]$ as a function of equivalent shape ratio for 5% damping in the basin.

TABLE CAPTIONS

Table 1. Mechanical properties for basin and halfspace. Properties with subscript 2 correspond to the basin, and properties with subscript 1 correspond to the halfspace.

Table 2. Amplification factors for first and second elastic modes of basin vibration due to a vertically incident P-wave.

Table 3. Amplification factors for first and second modes of damped basin vibration due to a vertically incident P-wave.

FIGURES

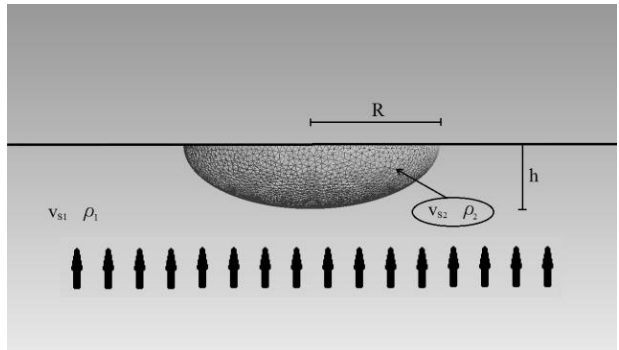

Figure 1. A 3D semi-cylindrical basin embedded in a half space and subjected to a plane wavefield.

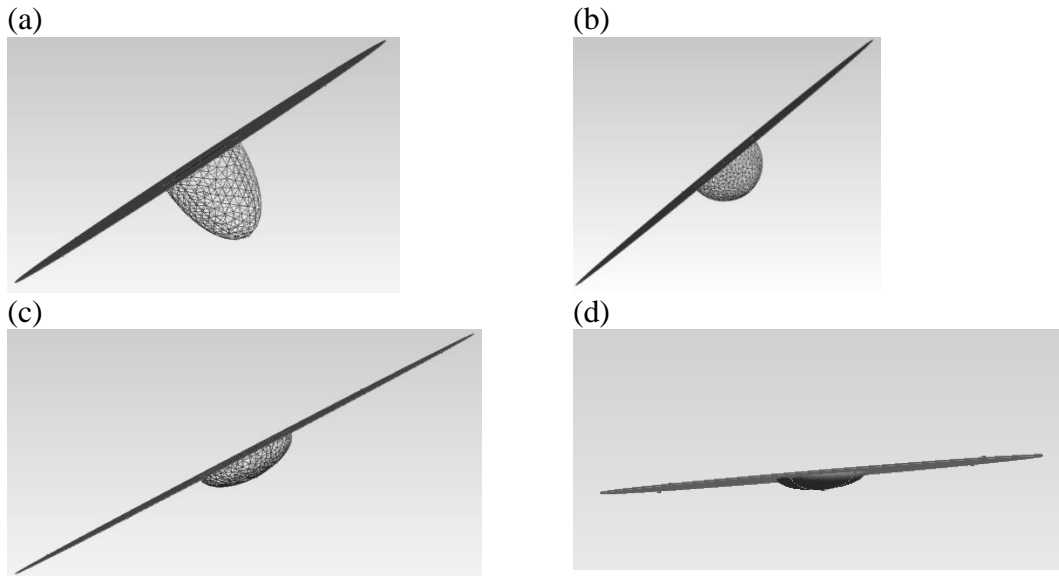

Figure 2. Basins with different aspect ratios $R/h$. (a) $R/h = 0.5$ (b) $R/h = 1$ (c) $R/h = 2$ (d) $R/h = 3$.

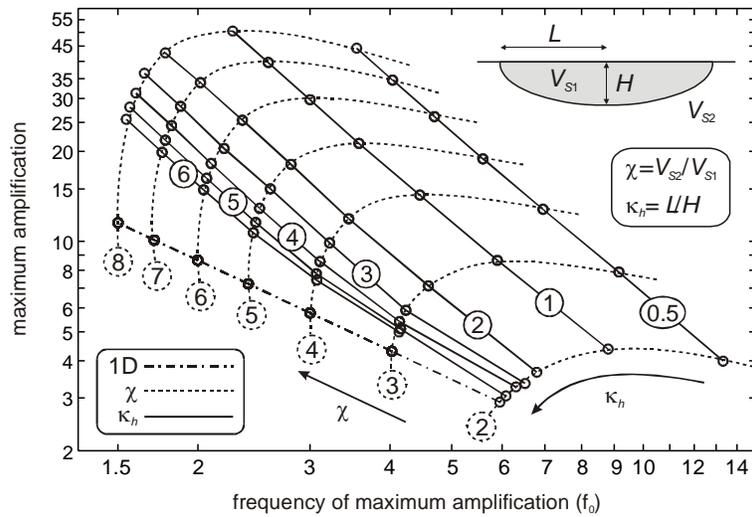

Figure 3. Maximum amplification and related frequencies (in Hz) for variable shape ratios $k_h$ and velocity ratios $\chi$. Here the symbol L has been used instead of R for the radius of the basin. Taken from Semblat *et al.* (2010).

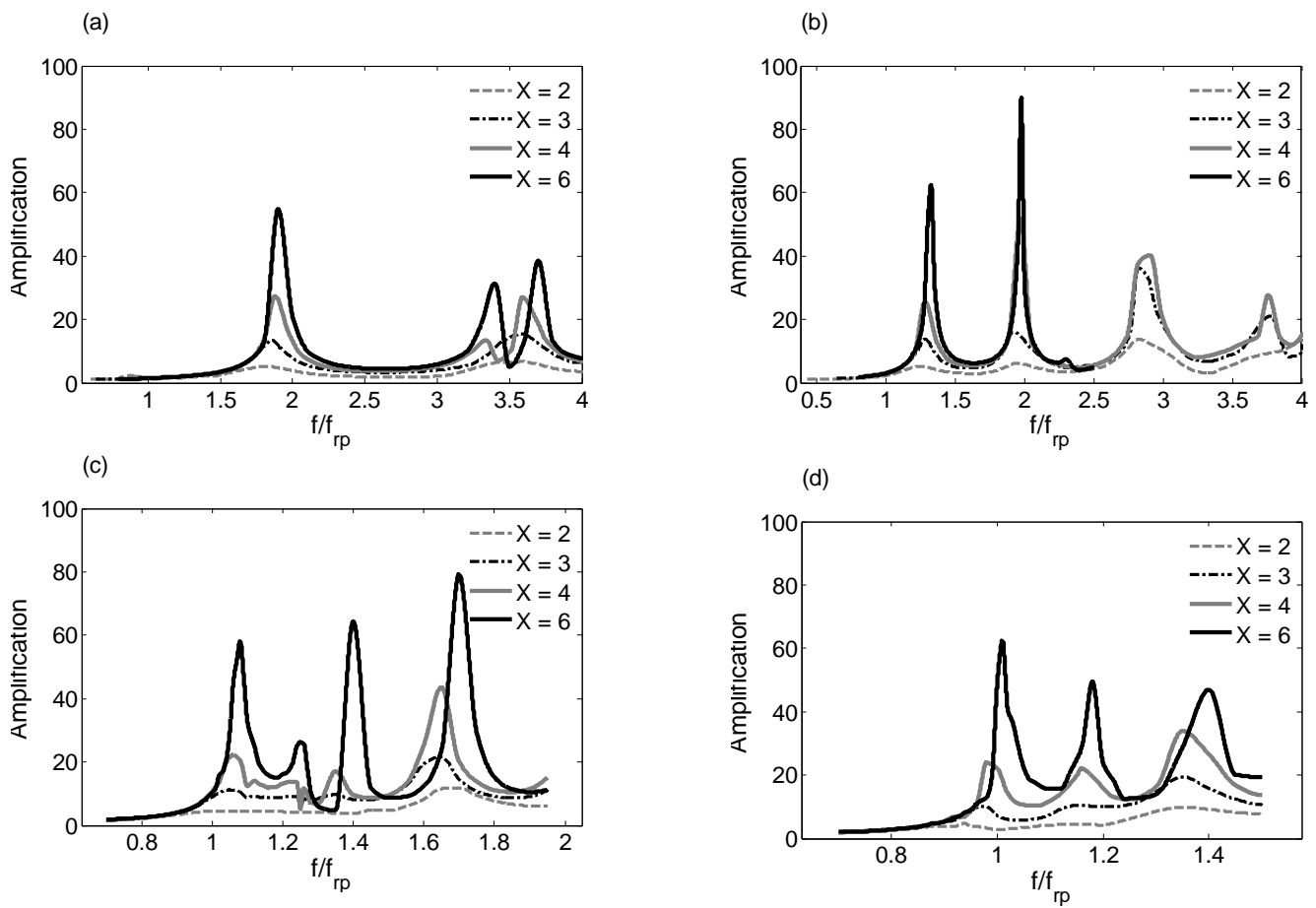

Figure 4. Amplification at the top of the undamped 3D basin due to vertically incident *P*-waves for different aspect ratios. (a) $R/h = 0.5$ (b) $R/h = 1$ (c) $R/h = 2$ (d) $R/h = 3$.

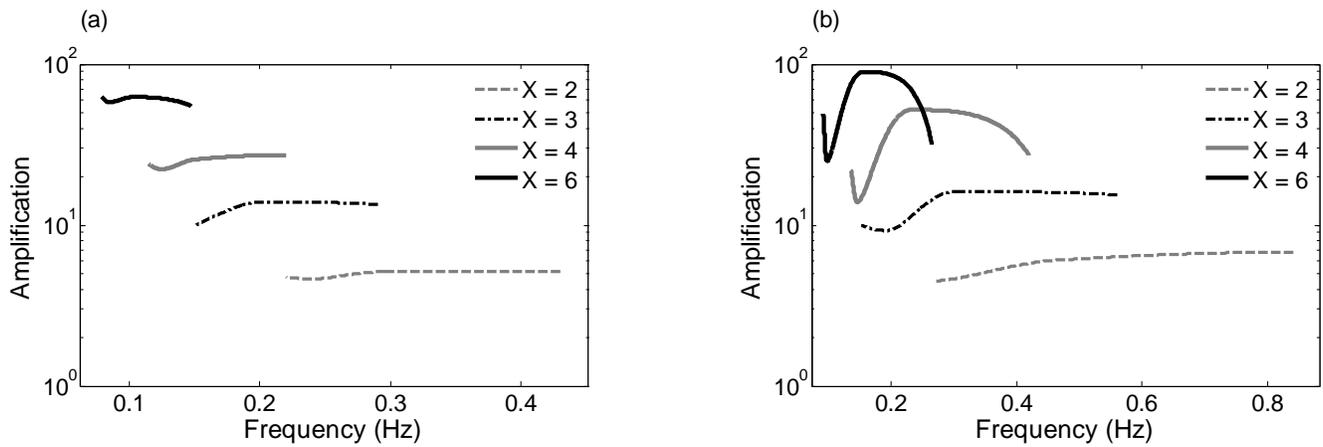

Figure 5. Amplification spectra for the first and second elastic modes of a semi-spherical basin for a vertically incident *P*-wave. (a) First mode (b) Second mode.

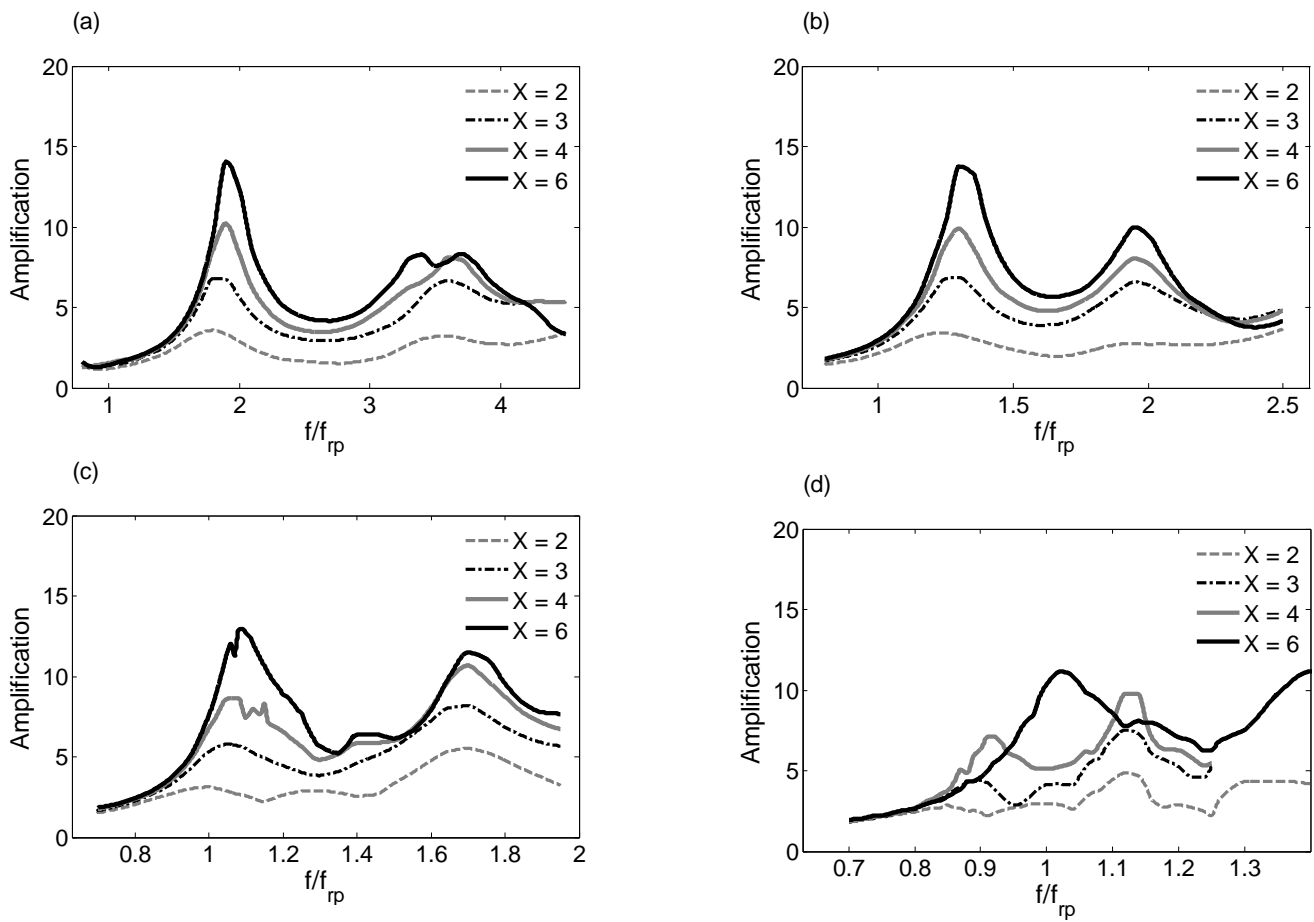

Figure 6. Amplification at the top of the 3D basin with 5% damping due to vertically incident *P*-waves. (a) $R/h = 0.5$ (b) $R/h = 1$

(c) $R/h = 2$ (d) $R/h = 3$.

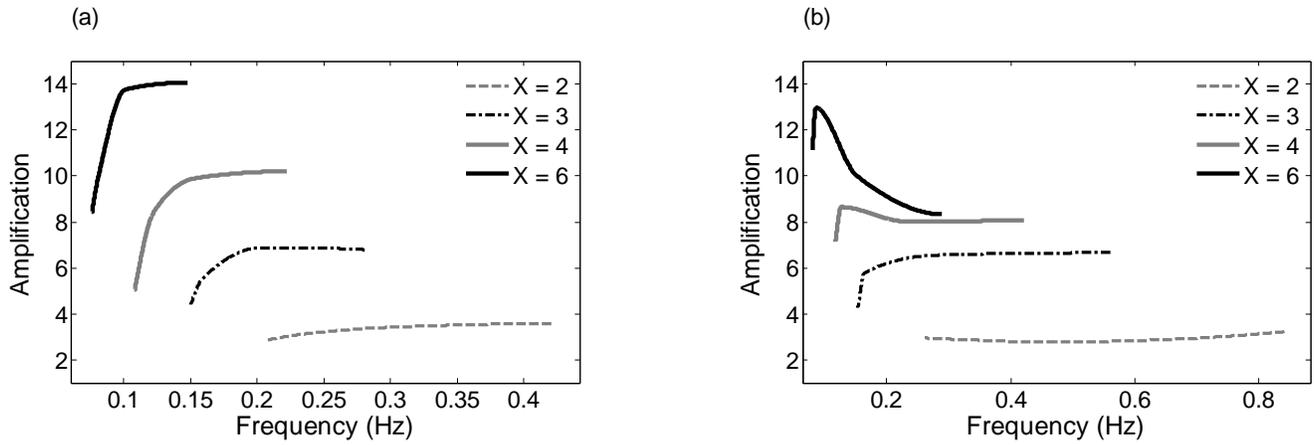

Figure 7. Amplification spectra for the first and second modes of 3D damped semi-spherical basin due to a vertically incident *P*-wave. (a) First mode (b) Second mode.

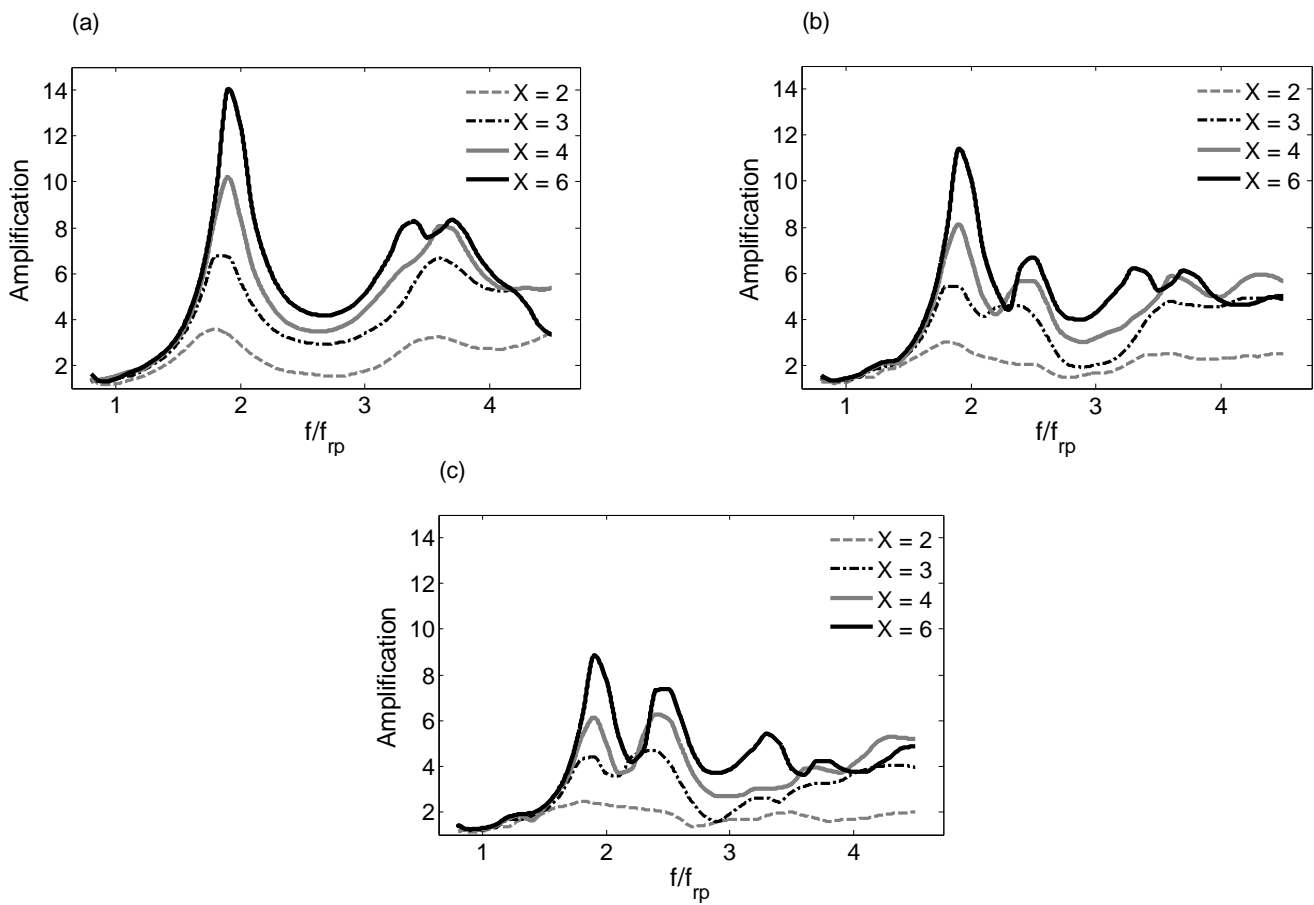

Figure 8. Amplification at the top of the 3D deepest basin with 5% damping and $R/h = 0.5$ due to incident *P*-waves. (a) Vertical incident (b) $\theta = 30$ degrees (c) $\theta = 45$ degrees

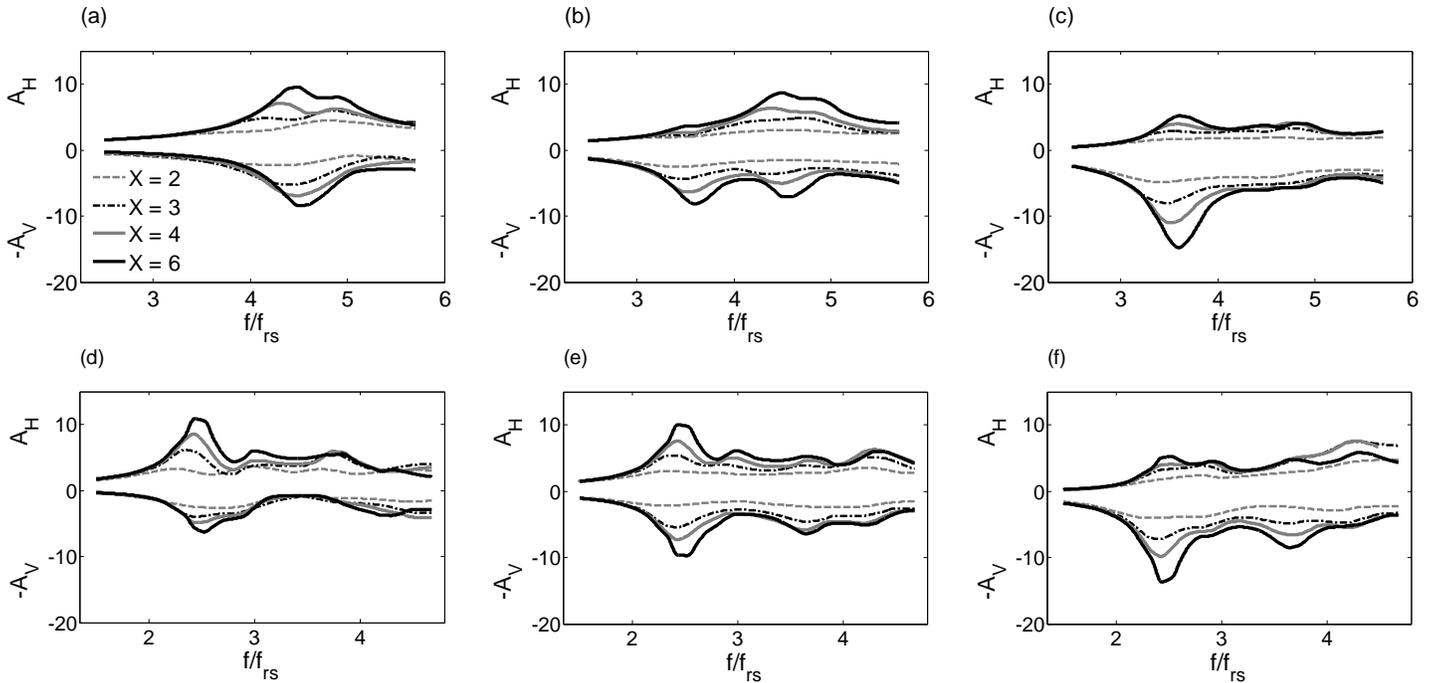

Figure 9. Comparison of amplification at the top of the 3D basin with 5% damping due to *S*-waves with different angles of incidence. (a) $R/h = 0.5$, $\theta = 0$ degrees (b) $R/h = 0.5$, $\theta = 30$ degrees (c) $R/h = 0.5$, $\theta = 45$ degrees (d) $R/h = 1$, $\theta = 0$ degrees (e) $R/h = 1$, $\theta = 30$ degrees (f) $R/h = 1$, $\theta = 45$ degrees.

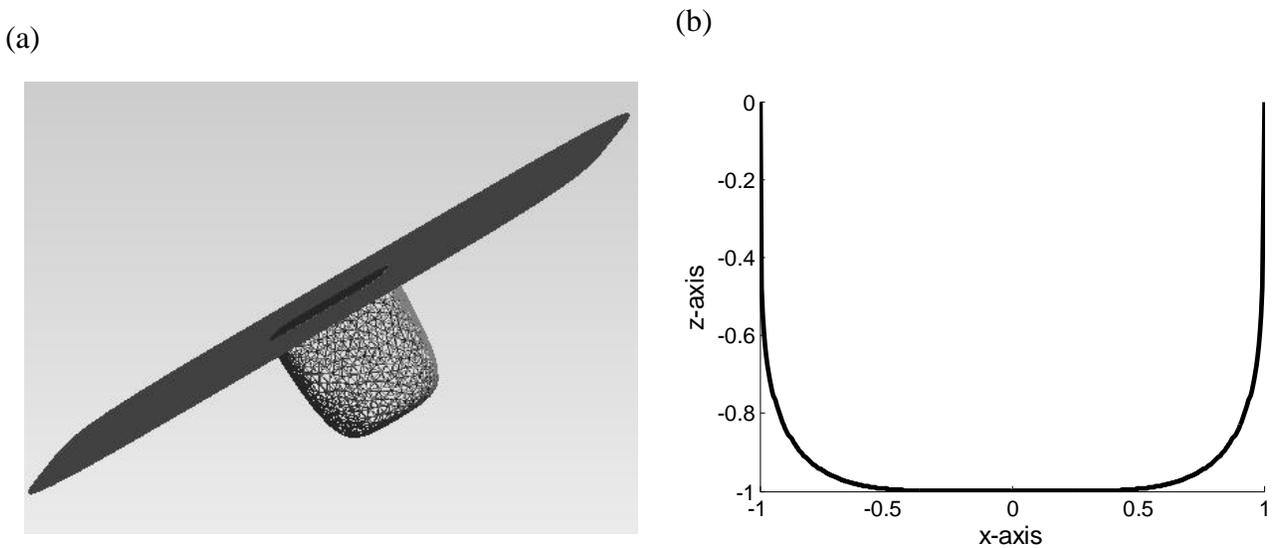

Figure 10. 3D basin model with super-ellipsoid shape. (a) View of the mesh for the super-ellipsoid and the truncated free surface. (b) A cross section corresponding to the *x-z* plane.

(a)                                      (b)

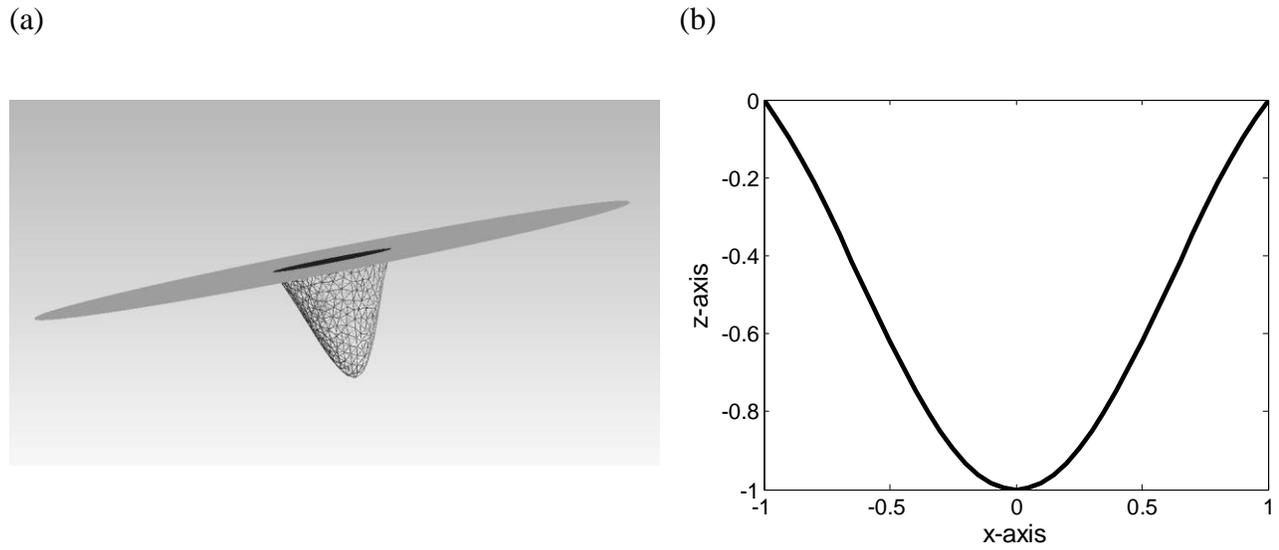

Figure 11. 3D basin model with shape of a 3D half sine. (a) View of the mesh for the 3D cosine and the truncated free surface. (b) A cross section corresponding to the $x$-$z$ plane (the period is adjusted so that the height of the bell is only 90% of that of a full cosine cycle).

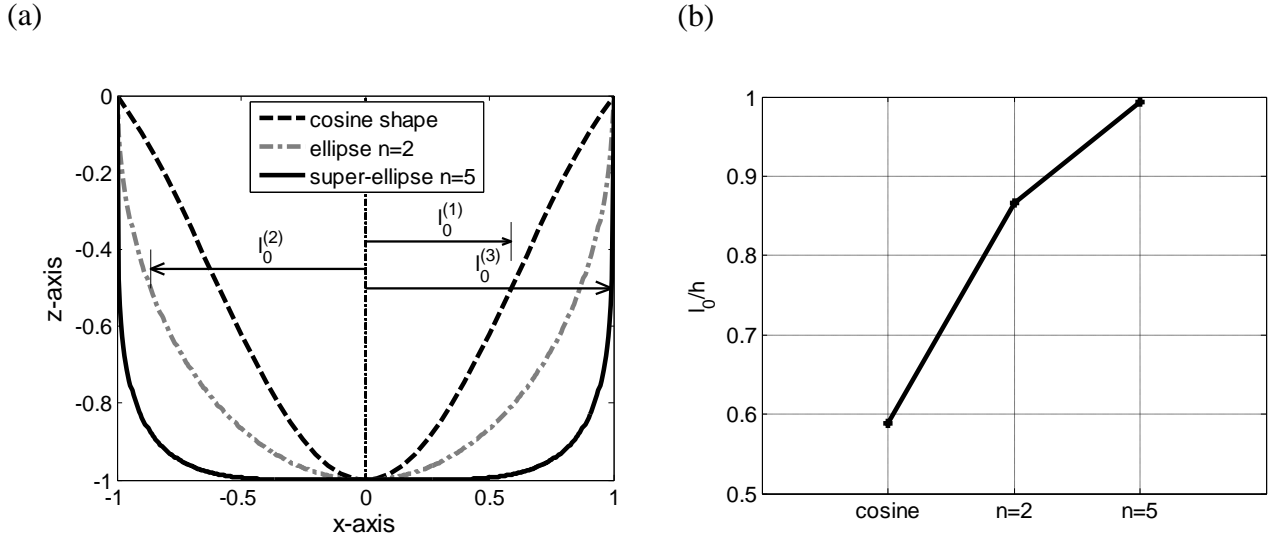

Figure 12. Basin shapes considered in the study. (a) Cross section corresponding to the $x$-$z$ plane. (b) Equivalent shape ratio $l_0/h$ for the three basin shapes for a unitary radius.

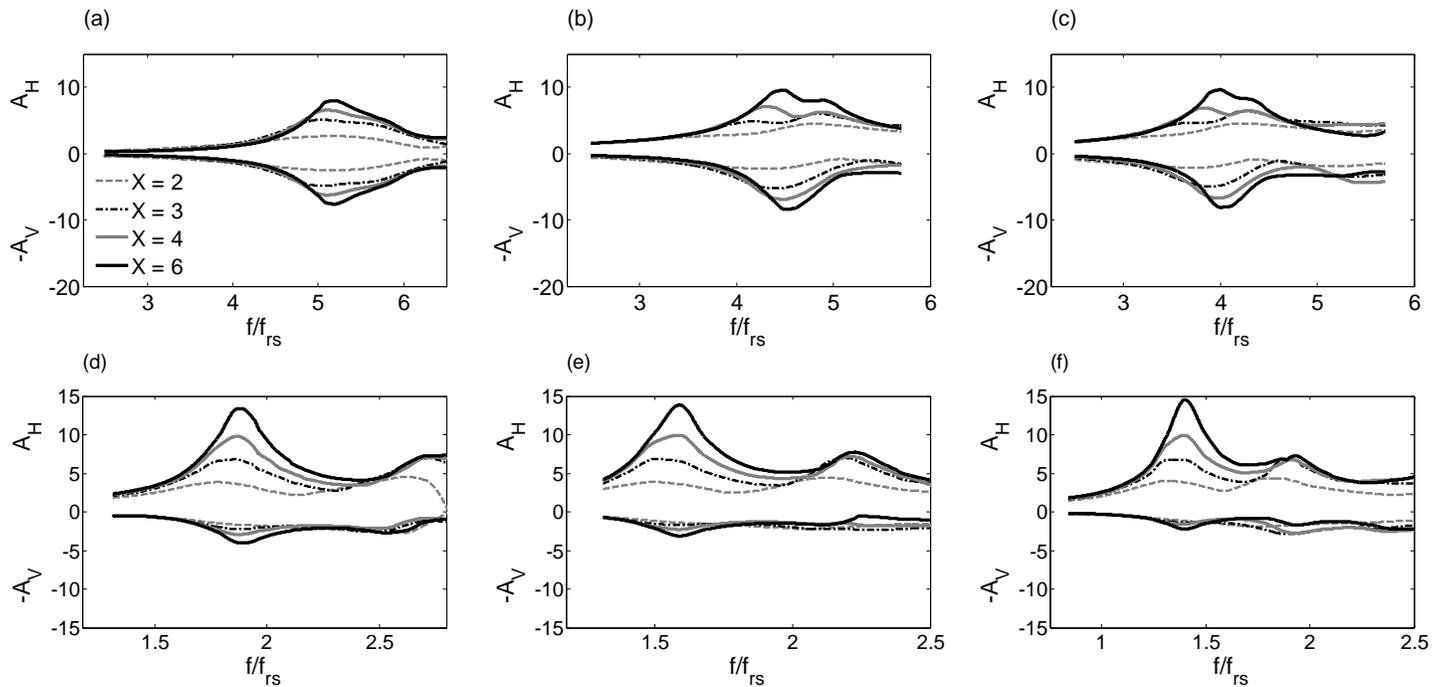

Figure 13. Amplification at the top of the 3D basin with 5% damping due to vertically incident $S$-waves with different equivalent shape ratios. (a) $R/h = 0.5$, basin with cosine shape (b) $R/h = 0.5$, basin with ellipsoidal shape, $n = 2$ (c) $R/h = 0.5$, basin with super-ellipsoidal shape, $n = 5$ (d) $R/h = 2$, basin with cosine shape (e) $R/h = 2$, basin with ellipsoidal shape, $n = 2$ (f) $R/h = 2$,

basin with super-ellipsoidal shape, $n = 5$.

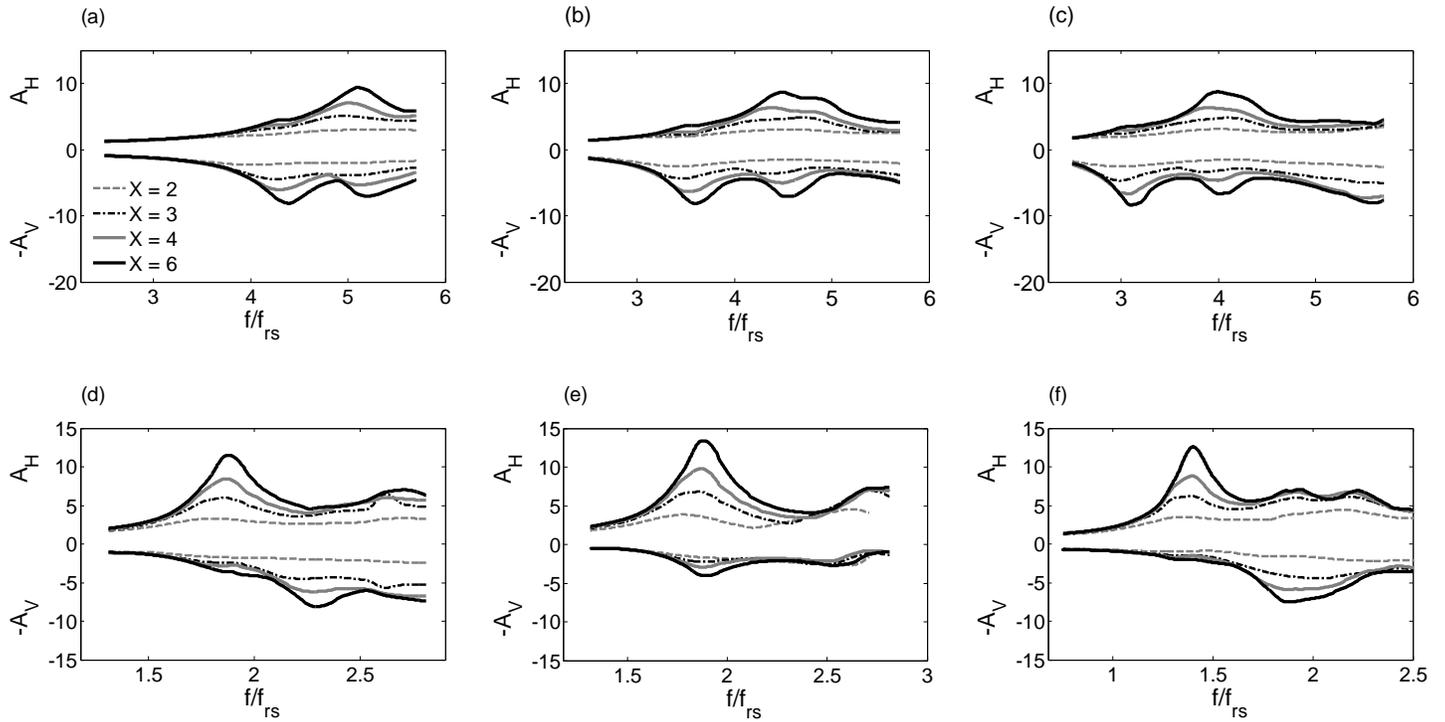

Figure 14. Amplification at the top of the 3D basin with 5% damping due to $S$-waves with incidence angle $\theta = 30°$. (a) $R/h = 0.5$, basin with cosine shape (b) $R/h = 0.5$, basin with ellipsoidal shape, $n = 2$ (c) $R/h = 0.5$, basin with super-ellipsoidal shape, $n = 5$ (d) $R/h = 2$, basin with cosine shape (e) $R/h = 2$, basin with ellipsoidal shape, $n = 2$ (f) $R/h = 2$, basin with super-ellipsoidal shape, $n = 5$.

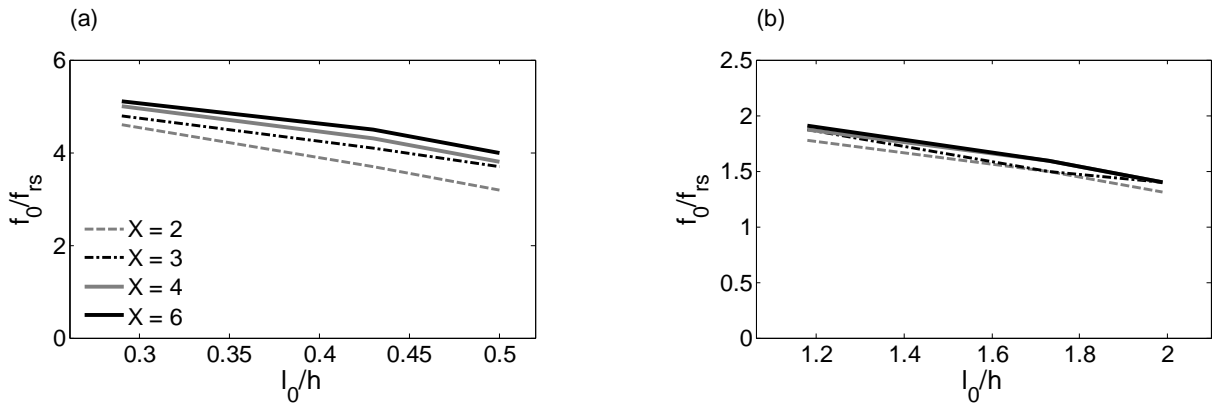

Figure 15. Normalized fundamental frequency $f_0/f_{rs}$ as a function of equivalent shape ratio $\tilde{k}_h$. (a) Basin with aspect ratio $\kappa_h = 0.5$ (b) Basin with aspect ratio $\kappa_h = 2$.

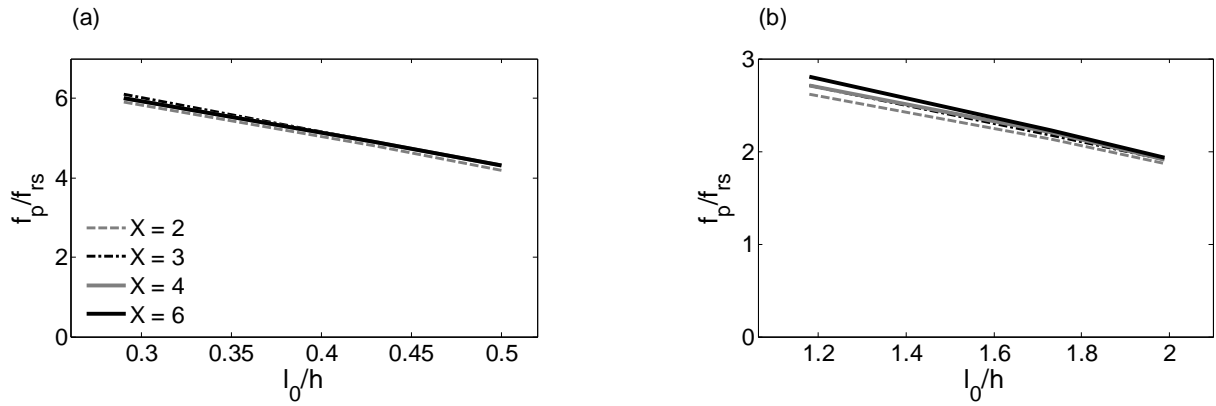

Figure 16. Normalized predominant frequency $f_p/f_{rs}$ as a function of equivalent shape ratio $\tilde{\kappa}_h$. (a) Basin with aspect ratio $\kappa_h = 0.5$ (b) Basin with aspect ratio $\kappa_h = 2$.

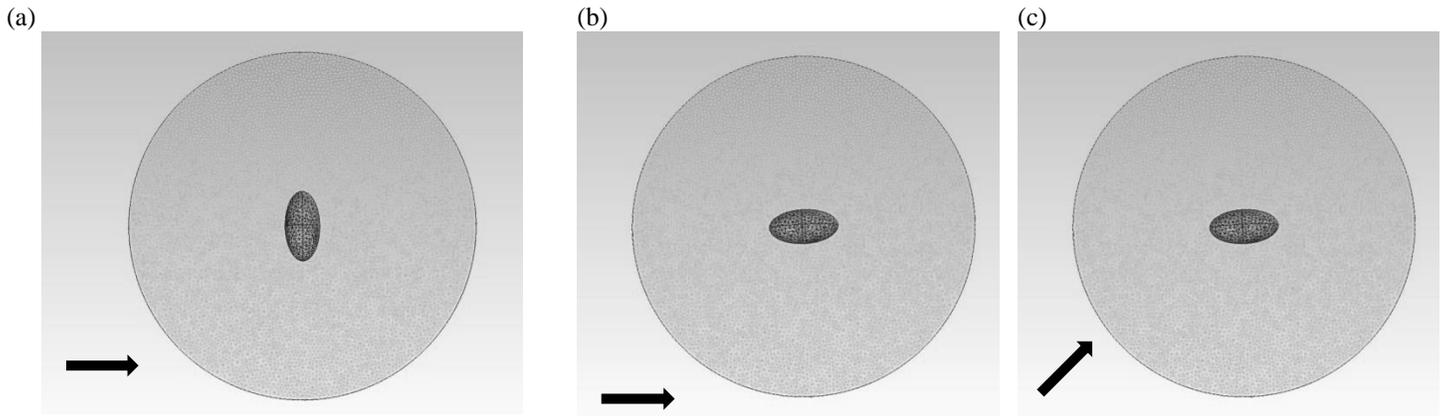

Figure 17. 3D basin models with asymmetry. (a) Basin with $R_x/R_y = 0.5$ (b) Basin with $R_x/R_y = 2$. (c) Basin with $R_x/R_y = 2$, direction of polarization of incident plane wave 45°.

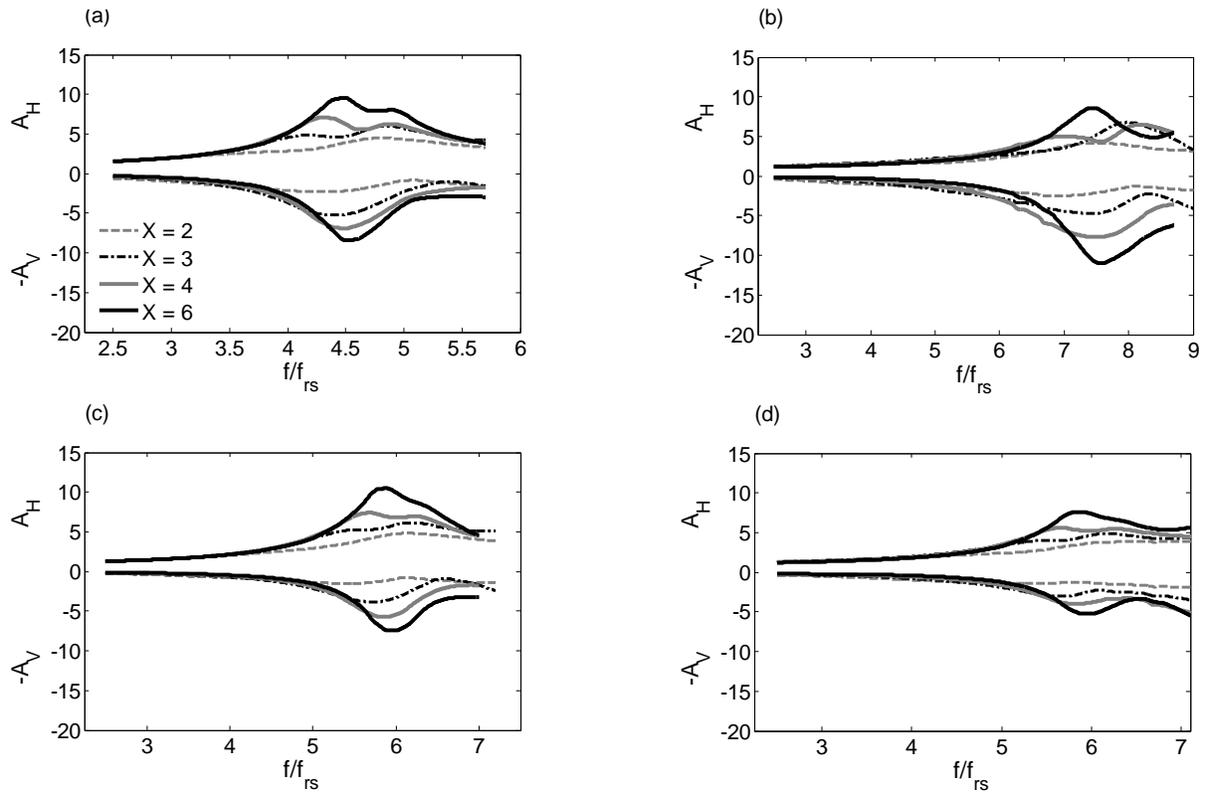

Figure 18. Amplification at the top of the 3D basin with 5% damping due to vertically incident *S*-waves for $R/h = 0.5$. (a) Symmetric basin (b) Basin with $\kappa_{xy} = 0.5$ (c) Basin with $\kappa_{xy} = 2$ (d) Basin with $\kappa_{xy} = 2$ and polarization direction at 45°.

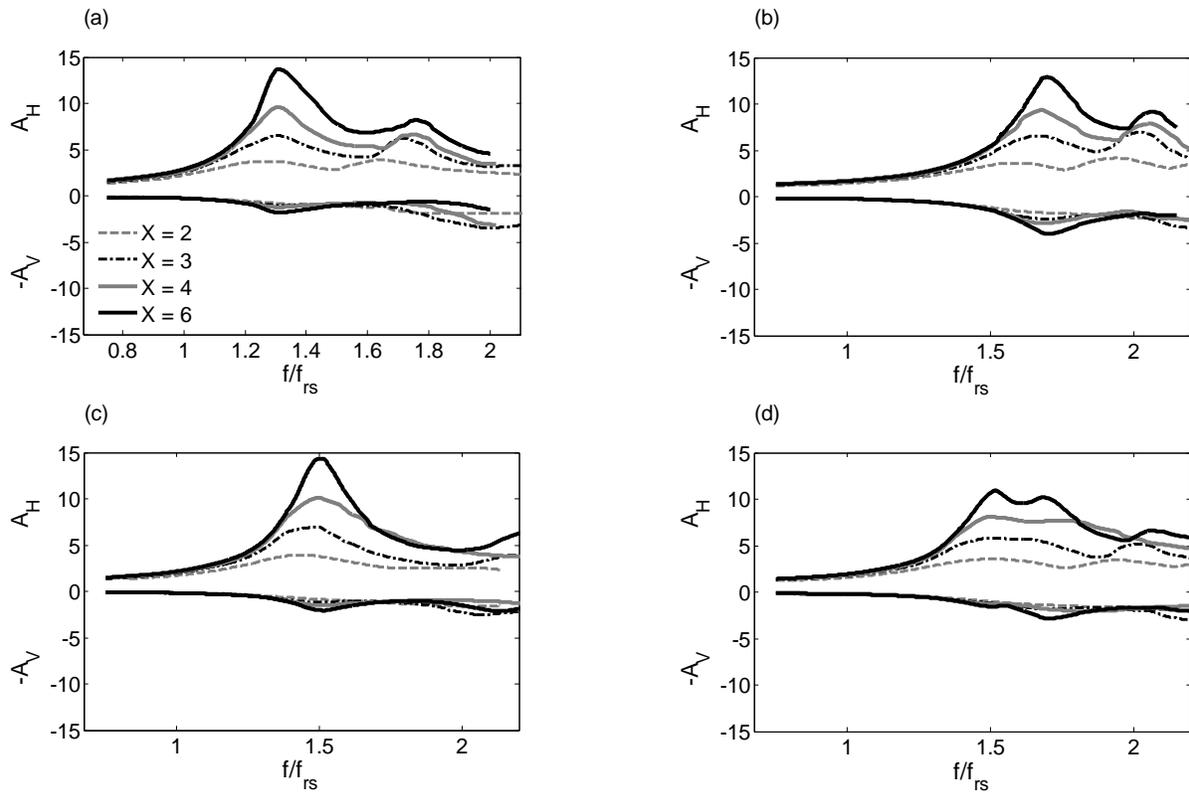

Figure 19. Amplification at the top of the 3D basin with 5% damping due to vertically incident *S*-waves for $R/h = 3$. (a) Symmetric basin (b) Basin with $\kappa_{xy} = 0.5$ (c) Basin with $\kappa_{xy} = 2$ (d) Basin with $\kappa_{xy} = 2$ and polarization direction at 45°.

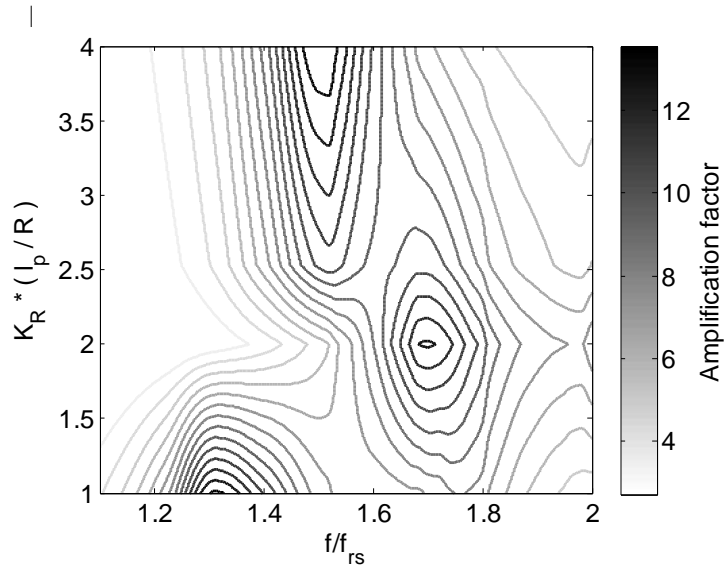

Figure 20. Amplification factor at the top of the basin with 5% damping due to vertically incident *S*-waves for 3D symmetric and asymmetric basins.

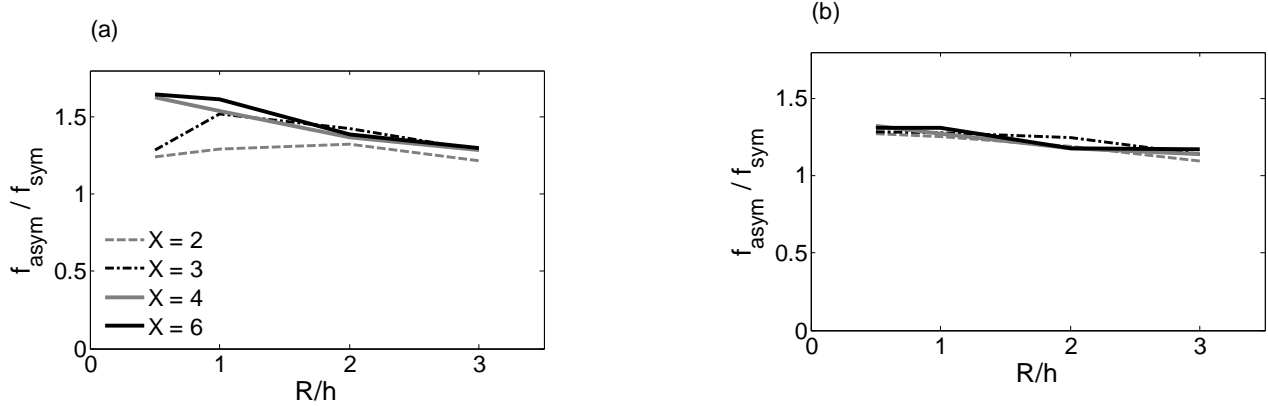

Figure 21. Fundamental frequency in a 3D asymmetric basin normalized with respect to the fundamental frequency of the symmetric basin. (a) Basin with $R_x/R_y = 0.5$ (b) Basin with $R_x/R_y = 2$.

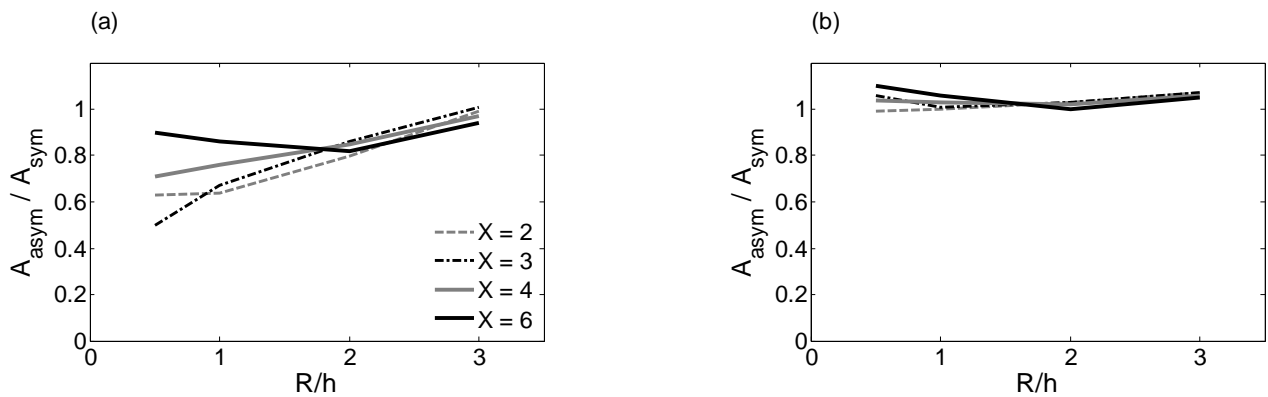

Figure 22. Amplification in a 3D asymmetric basin normalized with respect to the amplification of a symmetric basin. (a) Basin with $R_x/R_y = 0.5$ (b) Basin with $R_x/R_y = 2$.

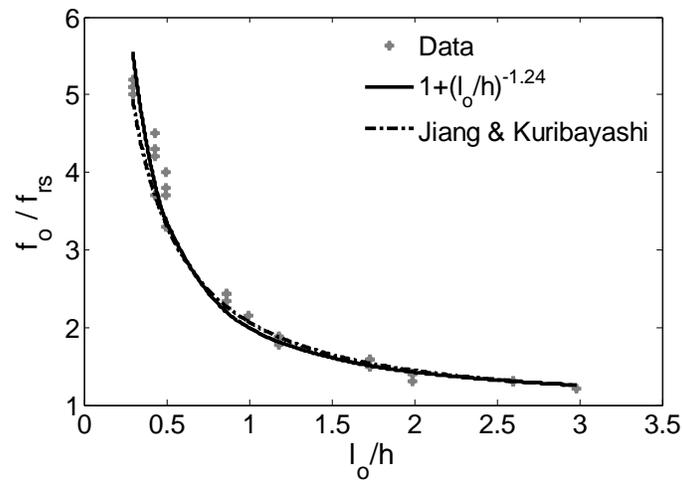

Figure 23. Normalized fundamental frequency $f_o/f_{rs}$ as a function of equivalent shape ratio.

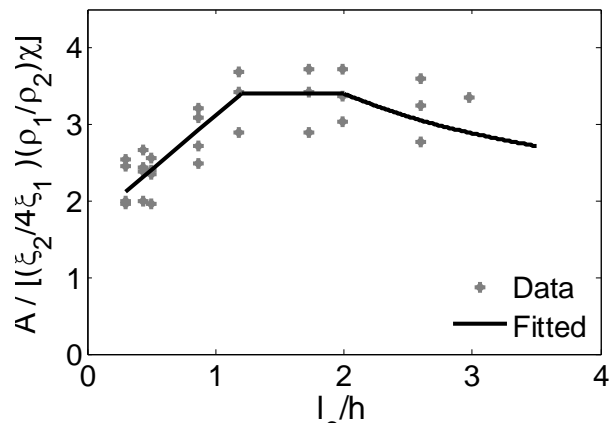

Figure 24. 3D/1D amplification factor $A/[(\xi_2/4\xi_1)(\rho_1/\rho_2)\chi]$ as a function of equivalent shape ratio for 5% damping in the basin.

TABLES

| Parameter | Value |
|---|---|
| Maximum depth ($h$) | 1 |
| Aspect ratio ($\kappa_h = R/h$) | $R$ |
| Velocity ratio ($\chi$) | $v_{s1}/v_{s2}$ |
| Ratio of densities ($\rho_2/\rho_1$) | 0.6 |
| Poisson ratio of basin ($\nu_2$) | 0.30 |
| Poisson ratio of halfspace ($\nu_1$) | 0.25 |

Table 1. Mechanical properties for basin and halfspace. Properties with subscript 2 correspond to the basin, and properties with subscript 1 correspond to the halfspace.

| $\chi = 2$ | | $\chi = 3$ | | $\chi = 4$ | | $\chi = 6$ | |
|---|---|---|---|---|---|---|---|
| fr (Hz) | A | fr (Hz) | A | fr (Hz) | A | fr (Hz) | A |
| First Mode | | | | | | | |
| 0.43 | 5.16 | 0.29 | 13.45 | 0.22 | 27.33 | 0.148 | 54.72 |
| 0.29 | 5.12 | 0.2 | 13.91 | 0.15 | 25.59 | 0.103 | 62.55 |
| 0.239 | 4.63 | 0.164 | 11.26 | 0.124 | 22.22 | 0.084 | 58.07 |
| 0.22 | 4.74 | 0.15 | 9.9 | 0.115 | 23.9 | 0.079 | 62.42 |
| Second Mode | | | | | | | |
| 0.84 | 6.8 | 0.56 | 15.45 | 0.42 | 27.06 | 0.265 | 31.42 |
| 0.46 | 6.04 | 0.3 | 16.16 | 0.23 | 52.04 | 0.154 | 89.79 |
| 0.276 | 4.51 | 0.193 | 9.25 | 0.145 | 13.9 | 0.097 | 25.03 |
| 0.269 | 4.49 | 0.153 | 9.99 | 0.136 | 22.08 | 0.092 | 49.34 |

Table 2. Amplification factors for first and second elastic modes of basin vibration due to a vertically incident *P*-wave.

| $\chi = 2$ | | $\chi = 3$ | | $\chi = 4$ | | $\chi = 6$ | |
|---|---|---|---|---|---|---|---|
| fr (Hz) | A | fr (Hz) | A | fr (Hz) | A | fr (Hz) | A |
| First Mode | | | | | | | |
| 0.421 | 3.58 | 0.281 | 6.79 | 0.222 | 10.2 | 0.148 | 14.06 |
| 0.292 | 3.4 | 0.203 | 6.88 | 0.152 | 9.89 | 0.101 | 13.77 |
| 0.234 | 3.12 | 0.159 | 5.56 | 0.122 | 8.4 | 0.08 | 9.77 |
| 0.206 | 2.86 | 0.15 | 4.4 | 0.108 | 5.04 | 0.076 | 8.45 |
| Second Mode | | | | | | | |
| 0.842 | 3.23 | 0.561 | 6.66 | 0.421 | 8.09 | 0.288 | 8.34 |
| 0.456 | 2.78 | 0.304 | 6.58 | 0.228 | 8.03 | 0.152 | 9.99 |
| 0.304 | 2.9 | 0.164 | 5.78 | 0.126 | 8.65 | 0.085 | 12.97 |

| 0.262 | 2.95 | 0.153 | 4.27 | 0.117 | 7.14 | 0.08 | 11.14 |

Table 3. Amplification factors for first and second modes of damped basin vibration due to a vertically incident *P*-wave.